\documentclass[twocolumn,prl]{revtex4-1}
\usepackage{graphicx}
\usepackage{amsmath,amssymb,bm,amsthm}
\usepackage{color}
\usepackage{here}
\usepackage{amsmath,amssymb,bm,amsthm,cases}

\usepackage{url}
\usepackage{natbib}
\usepackage[colorlinks=true,linkcolor=blue,citecolor=blue,breaklinks=true]{hyperref}

\usepackage{newtxtext,newtxmath}
\def\beq{\begin{equation}}
\def\eeq{\end{equation}}
\def\nbeq{\begin{equation*}}
\def\neeq{\end{equation*}}

\def\<{\langle}
\def\>{\rangle}

\usepackage{braket}
\newcommand{\sectionprl}[1]{{\par\textit{#1}.---}}

\newcommand{\nc}{\newcommand}
\nc{\rnc}{\renewcommand}
\nc{\nn}{\nonumber}

\begin{document}
\title{Construction of quantum dark soliton in one-dimensional Bose gas}

\author{Eriko Kaminishi}
\affiliation{Quantum Computing Center, Keio University, 3-14-1 Hiyoshi, Kohoku-ku, Yokohama}

\author{Takashi Mori}
\affiliation{RIKEN Center for Emergent Matter Science (CEMS), Wako 351-0198, Japan}

\author{Seiji Miyashita}
\affiliation{Department of Physics, University of Tokyo, 7-3-1 Hongo, Bunkyo-ku, Tokyo 113-0033, Japan}
\affiliation{Elements Strategy Initiative Center for Magnetic Materials (ESICMM), National Institute for Materials Science, Tsukuba, Ibaraki, Japan}

\date{\today}

\begin{abstract}
Dark soliton solutions in the one-dimensional classical nonlinear Schr\"odinger equation has been considered to be related to the yrast states corresponding to the type-II excitations in the Lieb-Liniger model.
However, the relation is nontrivial and remains unclear because a dark soliton localized in space breaks the translation symmetry, while yrast states are translationally invariant.
In this work, we construct a symmetry-broken quantum soliton state and investigate the relation to the yrast states.
By interpreting a quantum dark soliton as a Bose-Einstein condensation to the wave function of a classical dark soliton, we find that the quantum soliton state has a large weight only on the yrast states, which is analytically proved in the free-boson limit and numerically verified in the weak-coupling regime.
By extending these results, we derive a parameter-free expression of a quantum soliton state that is written as a superposition of yrast states with Gaussian weights.
The density profile of this quantum soliton state excellently agrees to that of the classical dark soliton.
The dynamics of a quantum dark soliton is also studied, and it turns out that the density profile of a dark soliton decays, but the decay time increases as the inverse of the coupling constant in the weak-coupling limit.  
\end{abstract}
\maketitle

\sectionprl{Introduction}
Ultracold bosonic atoms form a Bose-Einstein condensate (BEC).
All the particles in the condensate can be well described by a single wave function, which is called a ``macroscopic wave function''~\cite{London_text,Ruprecht1995}.
The macroscopic wave function obeys the classical nonlinear Schr\"odinger/ Gross-Pitaevskii equation~\cite{Pitaevskii1961,Gross1961}.
Such a nonlinear wave equation may possess solitary wave solutions whose shape do not change in the time evolution.
Indeed, in one dimension, the nonlinear Schr\"odinger equation possesses bright~\cite{Shabat1972} and dark~\cite{Tsuzuki1971} soliton solutions for attractive and repulsive interactions, respectively.

The one-dimensional classical nonlinear Schr\"odinger equation corresponds to the classical field approximation of the Lieb-Liniger model~\cite{Lieb1963}, which is a representative of quantum integrable models.
Thus, it has been desired to understand the relation between many-body energy eigenstates of the Lieb-Liniger Hamiltonian and the classical soliton solutions of the nonlinear Schr\"odinger equation.

The problem is that in the periodic boundary condition, an individual energy eigenstate has a flat single-particle probability density due to the translation invariance, and hence, we should consider some nontrivial superposition of energy eigenstates to construct a symmetry-broken state with a solitonic density profile.
As for the attractive interactions, such a quantum-classical correspondence has been established.
The attractive Lieb-Liniger model has bound states with momenta $\{P\}$, and it has been analytically shown that a quantum state corresponding to a classical bright soliton at position $X$ can be constructed by performing the Fourier transform of those bound states~\cite{Wadati-Sakagami1984}.
 
On the other hand, this problem has not been settled in a satisfactory manner for repulsive interactions.
It has been argued that a set of yrast states (energy eigenstates corresponding to Lieb's type-II excitations), each of which is the energy eigenstate with the lowest energy at a given momentum, in the Lieb-Liniger model is related to the family of classical dark solitons with momenta $P\in[-\pi\rho_0,\pi\rho_0]$, where $\rho_0$ is the particle number density~\cite{Ishikawa-Takayama1980, Kanamoto2009, Kanamoto2010}.
Indeed, the dispersion relation of yrast states is similar to that of the classical dark solitons in the weak-coupling regime~\cite{Ishikawa-Takayama1980}.
As pointed out above, however, a single energy eigenstate cannot represent a dark soliton at a fixed position.
It is not at all obvious how one can construct a many-body symmetry-broken state that corresponds to a classical dark soliton.

Sato et al.~\cite{Sato2012,Sato2016} tried to construct such a quantum many-body state guided by an analogy with the case of attractive interactions.
They considered the Fourier transform of yrast states $\ket{N,P}_\mathrm{yr}$ as an $N$-particle quantum dark soliton state at position $X$, i.e., $\ket{N,X}=\sum_Pe^{iP(X-L/2)}\ket{N,P}_\mathrm{yr}/\sqrt{N}$, where the sum is taken over $P=2\pi M/L$ with integer $M\in[0,N-1]$, and numerically found that the density profile in this state is similar to that in a classical dark soliton solution $\varphi_{P_0}(x-X)$ with a certain momentum $P_0$ at position $X$,
\begin{equation}
\bra{N,X}\hat{\psi}^\dagger(x)\hat{\psi}(x)\ket{N,X}\approx|\varphi_{P_0}(x-X)|^2.
\end{equation}
However, their construction is heuristic and there is no theoretical justification to consider the Fourier transform of $\ket{N,P}_\mathrm{yr}$.
More seriously, this method only reproduces a classical dark soliton at a certain momentum $P_0$.
It remains unclear how one can construct a quantum dark soliton state with a different momentum $P\neq P_0$, e.g., a completely ``black'' soliton with $P=\pi\rho_0$.

In this work, we study the quantum-classical correspondence of dark solitons, and construct a many-body quantum dark soliton state.
We start from the idea that a quantum soliton state should be interpreted as the Bose-Einstein condensation to the ``single-particle dark soliton state''.
It turns out that this BEC-like state is approximately expressed as a superposition of yrast states, which is concluded analytically in the free-boson limit, and checked numerically in a weak but finite coupling constant.
Based on this result, we propose a construction of the quantum dark soliton state by superposing yrast states.
It turns out that the quantum dark soliton is not obtained by the Fourier transform, but by the Gaussian superposition of type-II excitations.
The center of the Gaussian distribution determines the velocity of the dark soliton, and the width is found to be proportional to $c^{1/4}$ (see Eq.~(\ref{eq:P_variance}) below).
The density profile of the quantum dark soliton state constructed in this way shows an excellent agreement with that of the classical dark soliton solution.
Moreover, it turns out that this state has a lifetime longer than the soliton state constructed previously~\cite{Sato2012,Sato2016}.

\sectionprl{Setup}
We consider interacting bosons in a one-dimensional ring (i.e., the periodic boundary condition is imposed).
Such a system is described by the Lieb-Liniger model,
\begin{equation}
\hat{H}=\int_{-L/2}^{L/2}dx\left[-\hat{\psi}^\dagger(x)\partial_x^2\hat{\psi}(x)+c\hat{\psi}^\dagger(x)\hat{\psi}^\dagger(x)\hat{\psi}(x)\hat{\psi}(x)\right]
\end{equation}
with $c>0$ (i.e., repulsive interactions), where $\hat{\psi}(x)$ and $\hat{\psi}^\dagger(x)$ are annihilation and creation operators of a boson at $x$, which satisfies the commutation relations $[\hat{\psi}(x),\hat{\psi}^\dagger(y)]=\delta(x-y)$.
The number of particles is given by $N=\int_{-L/2}^{L/2}dx\,\hat{\psi}^\dagger(x)\hat{\psi}(x)$, and the average number density is denoted by $\rho_0=N/L$.

In the Lieb-Liniger model, the Bethe ansatz offers exact energy eigenstates and eigenvalues~\cite{Lieb1963}.
Each energy eigenstate $\ket{\{I_j\}_N}$ is characterized by a set of quantum numbers $I_1<I_2<\dots<I_N$, where $I_j$ is integer for odd $N$ and half-odd integer for even $N$.
For a given $\{I_j\}$, a set of quasi-momenta $k_1<k_2<\dots<k_N$ is obtained by the following Bethe ansatz equations:
\begin{equation}
k_j=\frac{2\pi}{L_j}I_j-\frac{2}{L}\sum_{l(\neq j)}^N\arctan\left(\frac{k_j-k_l}{c}\right).
\label{eq:BAE}
\end{equation}
The total momentum $P$ and the energy eigenvalue $E$ are obtained by $P=\sum_{j=1}^Nk_j$ and $E=\sum_{j=1}^Nk_j^2$.

The ground state corresponds to the quantum numbers $\{I_j\}=\{-(N-1)/2,-(N-1)/2+1,\dots,(N-1)/2\}$.
Yrast states are obtained by removing one of $I_j$ and adding $(N-1)/2+1$ (or $-(N-1)/2-1$), whose energy spectrum in a large finite system has been obtained~\cite{Kaminishi2011}.
It has been argued that yrast states correspond to the dark solitons since the dispersion relations coincide with each other in the weak-coupling limit after the thermodynamic limit~\cite{Ishikawa-Takayama1980}.

%The Heisenberg equation for the quantum field $\hat{\psi}(x)$ is given by
%\begin{equation}
%i\partial_t\hat{\psi}(x,t)=-\partial_x^2\hat{\psi}(x)+2c\hat{\psi}^\dagger(x)\hat{\psi}(x)\hat{\psi}(x).
%\end{equation}
The classical nonlinear Schr\"odinger equation is obtained as the classical limit of the Heisenberg equation for $\hat{\psi}(x)$; we replace the quantum field operator $\hat{\psi}(x)$ and $\hat{\psi}^\dagger(x)$ by the classical field $\psi(x)$ and $\psi^*(x)$, respectively, where $\psi^*$ is the complex conjugate of $\psi$:
\begin{equation}
i\partial_t\psi(x)=-\partial_x^2\psi(x)+2c\psi^*(x)\psi(x)\psi(x)-\mu\psi(x),
\label{eq:NLS}
\end{equation}
where $\mu$ is the chemical potential, which is given by $2\rho_0c$ in the thermodynamic limit, but there is a correction in a finite system~\cite{Sato2016}.
Equation~(\ref{eq:NLS}) has dark soliton solutions $\varphi_P(x-vt)$, where $P$ is the total momentum of a soliton and $v$ is the velocity that depends on $P$ through $v=dE/dP$.
It is noted that $|P|\leq\pi\rho_0$ and $|v|\leq |v_c|$, where $v_c$ is called the critical velocity.
In the thermodynamic limit, $v_c=2\sqrt{\rho_0c}$, but there is a correction in a finite-size system~\cite{Sato2016}.
The absolute square of $\varphi_P(x-x_0)$ corresponds to the particle number density of a dark soliton localized at $x_0$, and hence the normalization is given by $\int_{-L/2}^{L/2}dx\,|\varphi_P(x)|^2=N$.

An explicit expression of $\varphi_P(x)$ in a finite system is complicated but given in Ref.~\cite{Sato2016}.
In the thermodynamic limit, $\varphi_P^\infty(x)$ is given by
\begin{equation}
\varphi_P^\infty(x)=\sqrt{\rho_0}\left[\gamma\tanh\left(\gamma\sqrt{c\rho_0}x\right)+i\frac{v}{v_c}\right],
\end{equation}
where $\gamma=\sqrt{1-v^2/v_c^2}$ and $v$ is related to $P$ by the equation 
\begin{equation}
P(v)=2\rho_0\left\{\frac{\pi}{2}-\left[\frac{v}{v_c}\gamma+\arcsin\left(\frac{v}{v_c}\right)\right]\right\}
\label{eq:Pv}
\end{equation}
for $v\geq 0$, and $P(-v)=-P(v)$~\cite{Ishikawa-Takayama1980}.
In particular, the dark soliton at rest ($v=0$) corresponds to $P=\pi\rho_0$ and its wave function is given by
\begin{equation}
\varphi_{\pi\rho_0}^\infty(x)=\sqrt{\rho_0}\tanh(\sqrt{c\rho_0}x),
\end{equation}
which is completely black, i.e., $\varphi_{\pi\rho_0}^\infty(0)=0$.

\sectionprl{Quantum soliton state}
The success of the classical field approximation in BECs stems from the fact that it is a good picture that a macroscopically large number of particles occupy the same single-particle state with a wave function $\varphi(x)$.
If all the particles are in this state, the corresponding $N$-particle state is given by 
$$\frac{1}{\sqrt{N!}}\left(\int_{-L/2}^{L/2}dx\,\varphi(x)\hat{\psi}^\dagger(x)\right)^N\ket{\Omega},$$
where $\ket{\Omega}$ denotes the vacuum.
It is then natural to guess that an $N$-particle quantum soliton state corresponding to a classical dark soliton solution $\varphi_P(x-X)$ is given by the following BEC state:
\begin{equation}
\ket{N,X;P}=\frac{1}{\sqrt{N!}}\left(\int_{-L/2}^{L/2}dx\,\frac{\varphi_P(x-X)}{\sqrt{N}}\hat{\psi}^\dagger(x)\right)^N\ket{\Omega}.
\label{eq:Q_soliton}
\end{equation}
This quantum state has a desired property; it exactly reproduces the classical dark soliton density profile by
\begin{equation}
\bra{N,X;P}\hat{\psi}^\dagger(x)\hat{\psi}(x)\ket{N,X;P}_\mathrm{BEC}=|\varphi_P(x-X)|^2,
\end{equation}
as well as the wave function itself as
\begin{equation}
\hat{\psi}(x)\ket{N,X;P}=\varphi_P(x-X)\ket{N-1,X;P}.
\label{eq:soliton_wave}
\end{equation}

An important problem is to figure out the relation between $\ket{N,X;P}$ and the energy eigenstates $\ket{\{I_j\}_N}$ of the Lieb-Liniger model; we can always write
\begin{equation}
\ket{N,X;P}=\sum_{\{I_j\}}C_{\{I_j\}}\ket{\{I_j\}_N},
\end{equation}
but we want to understand the structure of expansion coefficients $C_{\{I_j\}}=\braket{\{I_j\}_N|N,X;P}$.

\begin{figure}[t]
\begin{center}
\includegraphics[width=8cm]{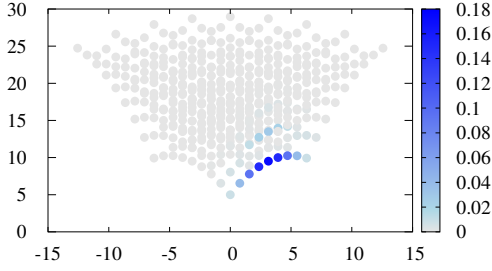}
\caption{The overlap $|C_{\{I_j\}}|^2=|\braket{\{I_j\}_N|N,X;P}|^2$ for $\rho_0=1$, $c=0.1$, and $N=8$.
Each circle represents an eigenstate $\{I_j\}$ with the momentum $P=\sum_{j=1}^Nk_j$ and the energy $E=\sum_{i=1}^Nk_j^2$.
The eigenstate with the lowest energy for a given momentum $P$ corresponds to an yrast state $\ket{N,P}_\mathrm{yr}$.}
\label{fig:overlap}
\end{center}
\end{figure}

First, we numerically calculate overlaps by using the determinant formulas for form factors~\cite{Gaudin1983,Korepin1982,Slavnov1989,Kojima1997,Caux2007}.
For simplicity, we focus on the quantum soliton state with $P=\pi\rho_0$.
Overlaps for $N=L=8$ ($\rho_0=1$) and $c=0.1$ are presented in Fig.~\ref{fig:overlap}.
Since the yrast state $\ket{N,P}_\mathrm{yr}$ with the momentum $P$ corresponds to the eigenstate with the lowest energy eigenvalue for the fixed momentum $P$, Fig.~\ref{fig:overlap} shows that the quantum soliton state has weights concentrated on yrast states.

%%%%%%%%%%%%%%%%%%%%% figure 1%%%%%%%%%%%%%%%%%%%%%%%%%%
%%%%%%%%%%%%%%%%%%%%%%%%%%%%%%%%%%%%%%%%%%%%%%%%%

Next, we analytically calculate the expansion coefficients in the free-boson limit, i.e., $c\to 0$ at a fixed $L$~\footnote
{It should be noted that the free-boson limit is different from the weak-coupling limit $c\to 0$ \textit{after} the thermodynamic limit.}.
In this limit, classical dark soliton solutions $\{\varphi_P(x)\}$ with $0\leq P\leq\pi\rho_0$ become
\begin{equation}
\varphi_P^\mathrm{free}(x)=\sqrt{\rho_0}\left(\sqrt{1-\frac{P}{2\pi\rho_0}}-\sqrt{\frac{P}{2\pi\rho_0}}e^{i\frac{2\pi}{L}x}\right),
\label{eq:soliton_free}
\end{equation}
which can be checked by taking the free-boson limit of the explicit expression of $\varphi_P(x)$ given in Ref.~\cite{Sato2016}.
On the other hand, the free-boson limit of yrast states $\ket{N,P}_\mathrm{yr}$ yields
\begin{equation}
\ket{N,P}_\mathrm{yr}^\mathrm{free}=\ket{n_0=N-M,n_{2\pi/L}=M},
\label{eq:II_free}
\end{equation}
where $M$ is an integer given by $P=2\pi M/L$, and the right-hand side means the state in which $N-M$ particles occupy the mode with the wave number $k=0$ and $M$ particles occupy the mode with $k=2\pi/L$~\footnote
{This state is indeed the eigenstate with the lowest energy for a given momentum $P$, and is therefore interpreted as an yrast state.}.
By using Eqs.~(\ref{eq:Q_soliton}), (\ref{eq:soliton_free}), and (\ref{eq:II_free}), we can analytically calculate the expansion coefficients.
The result is that expansion coefficients are nonzero only for yrast eigenstates $\{\ket{N,P}_\mathrm{yr}^\mathrm{free}\}$, and the quantum soliton state is expanded solely by yrast states as
\begin{equation}
\ket{N,X;P}^\mathrm{free}=\sum_{P'}e^{iP'(X-L/2)}e^{-\frac{N}{2}\varg_P^\mathrm{free}(P')}\ket{N,P'}_\mathrm{yr}^\mathrm{free},
\label{eq:Q_soliton_free}
\end{equation}
where the sum over $P'$ is taken for $P'=2\pi M/L$ with $M=0,1,\dots,N$.
The real function $\varg_P^\mathrm{free}(P')$ is given for large $N$ by
\begin{equation}
\varg_P^\mathrm{free}(P')\approx\left(1-\frac{P'}{2\pi\rho_0}\right)\ln\frac{2\pi\rho_0'-P'}{2\pi\rho_0-P}+\frac{P'}{2\pi\rho_0}\ln\frac{P'}{P}.
\end{equation}
It takes the minimum value at $P'=P$ and is expanded as
\begin{equation}
\varg_P^\mathrm{free}(P')\approx-\frac{(P'-P)^2}{2\sigma_\mathrm{free}^2}
\end{equation}
with 
\begin{equation}
\sigma_\mathrm{free}^2=\frac{1}{N}\frac{2\pi\rho_0}{(2\pi\rho_0)^{-1}+P^{-1}}\propto\frac{1}{N}.
\end{equation}
The soliton BEC state $\ket{N,X;P}$ is therefore expressed as a superposition of yrast states $\ket{N,P'}_\mathrm{yr}^\mathrm{free}$ with a Gaussian weight of mean $P$ and variance $\sigma_P^2$.
The quantum dark soliton constructed in the previous work~\cite{Sato2012} corresponds to the uniform superposition, i.e., $\varg_P(P')=\mathrm{const.}$, but this simple calculation in the free boson limit indicates the importance of the Gaussian weight for constructing a quantum soliton state.

\sectionprl{Gaussian superposition of yrast states}
From the numerical result for $c>0$ and the analytical result for the free-boson limit, it is expected that a quantum soliton state $\ket{N,X;P}$ is generically expressed as a superposition of yrast states.
By \textit{assuming} it, we can find how yrast states should be superposed to construct a quantum soliton state.
In the thermodynamic limit with a fixed $c$, the mean and the variance of the total momentum operator $\hat{P}=\int_{-L/2}^{L/2}dx\,\hat{\psi}^\dagger(x)(-i\partial_x)\hat{\psi}(x)$ are given by
\begin{equation}
\lim_{N\to\infty}\braket{N,X;P|\hat{P}|N,X;P}=P
\end{equation}
and
\begin{equation}
\lim_{N\to\infty}\braket{N,X;P|(\hat{P}-P)^2|N,X;P}=\frac{4}{3}\gamma^3\rho_0\sqrt{\rho_0c}\equiv\sigma_P^2,
\label{eq:P_variance}
\end{equation}
respectively.
In the BEC state $\ket{N,X;P}$, the momenta $\{p_j\}_{j=1}^N$ of $N$ particles can be considered to be independent random variables, and thus the central limit theorem implies that the total momentum $\sum_{j=1}^Np_j$ has a Gaussian distribution.
Therefore, if the quantum soliton state consists of the yrast states, the former is given by a Gaussian superposition of the latter for large system sizes:
\begin{align}
\ket{N,X;P}&\approx\mathcal{N}^{-1/2}\sum_{P'}e^{iP'(X-L/2)}e^{-(P'-P)^2/(2\sigma_P^2)}\ket{N,P'}_\mathrm{yr}
\nonumber \\
&\equiv\ket{N,X;P}_\mathrm{yr},
\label{eq:Q_soliton_yrast}
\end{align}
where $\mathcal{N}$ is a normalization constant, and the sum is taken over $P'=2\pi M/L$ with $M=0,1,\dots,N$.

\begin{figure}[t]
\begin{center}
\includegraphics[width=7cm]{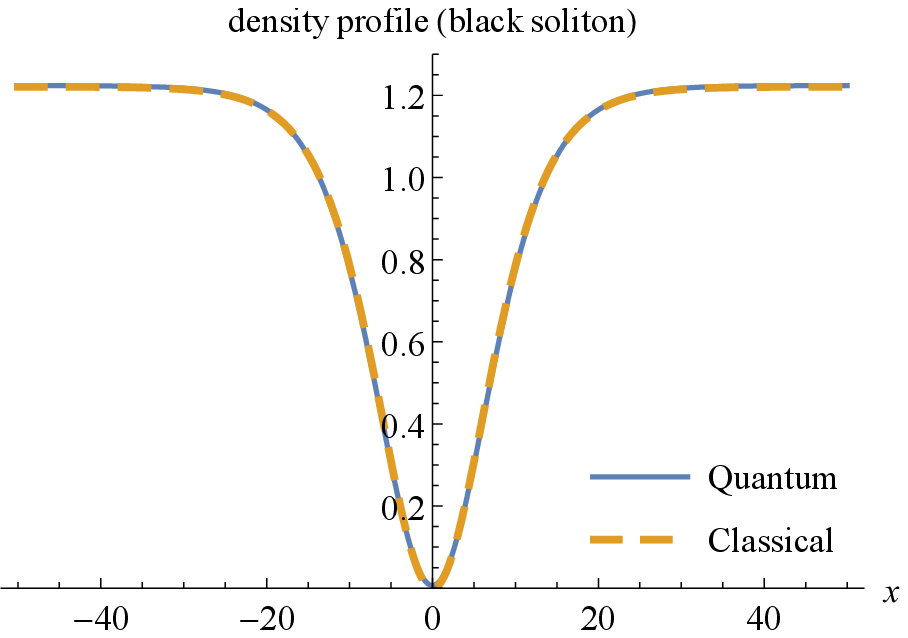}\\
\bigskip
\includegraphics[width=7cm]{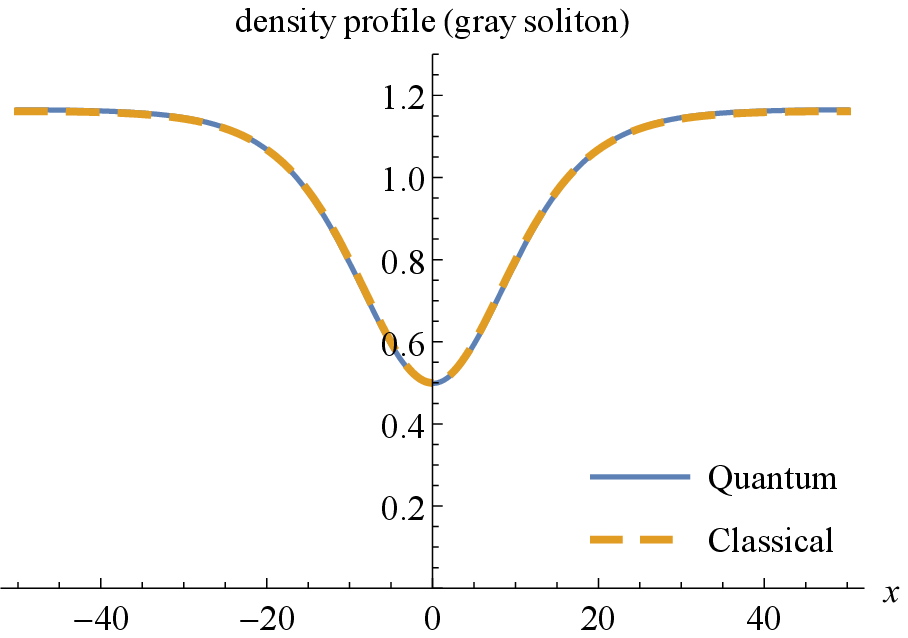}
\caption{Comparison of the density profiles of quantum soliton states $\ket{N,0;P}_\mathrm{yr}$ given by Eq.~(\ref{eq:Q_soliton_yrast}) and those of classical dark solitons for $P=\pi\rho_0$ (top) and $P=(\pi/2-1)\rho_0$ (bottom). The system size is set as $N=L=100$.}
\label{fig:density}
\end{center}
\end{figure}

Equation.~(\ref{eq:Q_soliton_yrast}) is greatly simplified compared to Eq.~(\ref{eq:Q_soliton}) since the former is restricted to the yrast states.
It is therefore possible to calculate some observables using the Bethe ansatz method for large system sizes.
In Fig.~\ref{fig:density}, we compare the density profile in the quantum soliton state of Eq.~(\ref{eq:Q_soliton_yrast}), $\braket{N,0,P|\hat{\psi}^\dagger(x)\hat{\psi}(x)|N,0,P}$ with its classical counterpart $|\varphi_P(x)|^2$ for $P=\pi\rho_0$ (black soliton) and $P=(\pi/2-1)\rho_0$ (gray soliton) for $N=L=100$ and $c=0.01$.
Quantum and classical solitons excellently agree with each other.

%%%%%%%%%%%%%%%%%%%%% figure 2%%%%%%%%%%%%%%%%%%%%%%%%%%
%%%%%%%%%%%%%%%%%%%%%%%%%%%%%%%%%%%%%%%%%%%%%%%%%

%By using Eq.~(\ref{eq:Q_soliton_yrast}), a single yrast state $\ket{N,P}_\mathrm{yr}$ can be expressed as a superposition of quantum soliton states as
%\begin{equation}
%\ket{N,P'}_\mathrm{yr}\propto\int_{-L/2}^{L/2}dx\,e^{-iP'x}\ket{N,x,P}.
%\end{equation}

%%%%%%%%%%%%%%%%%%%%% figure 3%%%%%%%%%%%%%%%%%%%%%%%%%%
\begin{figure}[t]
\begin{center}
\includegraphics[width=8cm]{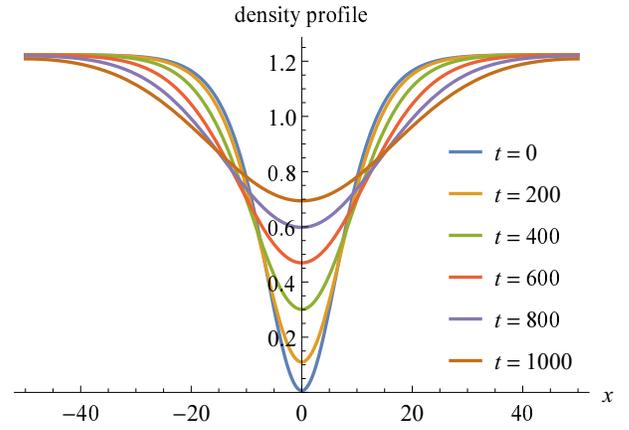}
\caption{Dynamics of the density profile starting from a quantum soliton state $\ket{N,0;\pi\rho_0}_\mathrm{yr}$ of Eq.~(\ref{eq:Q_soliton_yrast}) for $N=L=100$ and $c=0.01$.}
\label{fig:dynamics}
\end{center}
\end{figure}
%%%%%%%%%%%%%%%%%%%%%%%%%%%%%%%%%%%%%%%%%%%%%%%%%

\sectionprl{Time evolution}
Since energy eigenvalues are obtained by solving Eq.~(\ref{eq:BAE}) and using $E=\sum_{j=1}^Nk_j^2$, we can compute the dynamics of the density profile in a numerically exact manner. 
Figure~\ref{fig:dynamics} shows the time evolution of the expectation value of $\hat{\psi}^\dagger(x-vt)\hat{\psi}(x-vt)$, i.e., the density in the moving frame at the soliton velocity, starting from a quantum soliton state $\ket{N,0;\pi\rho_0}_\mathrm{yr}$, whose velocity is given by $v=2\pi/L$.
Since the quantum soliton state is not an eigenstate of the Hamiltonian, the solitonic density profile collapses.
The timescale $\tau$ of the collapse is evaluated by the Mandelstam-Tamm quantum speed limit~\cite{Mandelstam1945} as $\tau\geq\pi/(2\Delta E)$, where $\Delta E^2$ is the variance of the energy in an initial state.
In the quantum soliton state of Eq.~(\ref{eq:Q_soliton_yrast}), $\Delta E$ is given by
\begin{equation}
\Delta E\approx\left|\frac{\partial E}{\partial P}\Delta P+\frac{1}{2}\frac{\partial^2 E}{\partial P^2}\Delta P^2\right|
=\left|v\Delta P+\frac{1}{2}\frac{\partial v}{\partial P}\Delta P^2\right|.
\end{equation}
In the moving frame at the soliton velocity $v$, the first term $v\Delta P$ vanishes, so we have $\Delta E\approx|(\Delta P^2/2)\partial v/\partial P|$.
By using Eqs.~(\ref{eq:Pv}) and (\ref{eq:P_variance}) we obtain $\Delta E\approx\gamma^2\rho_0c/3$, and hence $\tau\approx 3\pi/(2\gamma^2\rho_0c)$.
The decay time of a quantum dark soliton is inversely proportional to $c$ in the weak-coupling regime, $\tau\sim 1/c$, which is confirmed numerically and consistent with Ref.~\cite{Sato2016}.

It is noted that the dependence of $\tau\sim 1/c$ has been reported by Sato et al.~\cite{Sato2016}, but quantitatively, the decay time of our quantum soliton state is much longer than that of the dark soliton state constructed in Ref.~\cite{Sato2016}.
Following Ref.~\cite{Sato2016}, if the decay time is defined by the time when the smallest value of the density notch reaches the value of 0.5, the decay time for $c=0.01$ is about 600 in our dark soliton state, while it is about 100 in the dark soliton state proposed in Ref.~\cite{Sato2016}.

\sectionprl{Conclusion and Discussion}
In this work, we have discussed the property of a quantum soliton state given by Eq.~(\ref{eq:Q_soliton}), which is interpreted as a Bose-Einstein condensation to a single-particle wave function $\varphi_P(x-x_0)/\sqrt{N}$, where $\varphi_P(x)$ is the dark soliton solution of the classical nonlinear Schr\"odinger equation.
It has turned out that this quantum soliton state almost consists of the yrast states, which is confirmed numerically for small $c>0$ and analytically for $c=0$.
This result offers a direct confirmation that a classical dark soliton with a localized position corresponds to a superposition of yrast states of the Lieb-Liniger model.
In addition, we have revealed that a quantum soliton state $\ket{N,X;P}$ is well approximated by the state $\ket{N,X;P}_\mathrm{yr}$ that has a Gaussian weight on each yrast state $\ket{N,P'}_\mathrm{yr}$ with mean $\braket{P'}=P$ and variance $\sigma_P^2=4\gamma^3\rho_0\sqrt{\rho_0c}/3$ in the weak-coupling regime.
Numerical calculations show excellent agreements between the density profile of the Gaussian superposition of the yrast states and that of the classical dark soliton.
We have also discussed dynamics of a quantum dark soliton, and it has been shown that the decay time is proportional to $c^{-1}$.

A remaining open problem is to understand the relation between the quantum soliton state constructed in this work, i.e., Eq.~(\ref{eq:Q_soliton}) or Eq.~(\ref{eq:Q_soliton_yrast}), and the recent theoretical observation by Syrwid and Sacha~\cite{Syrwid2015} that a dark soliton emerges in successive measurements of particle positions starting from a single yrast state.
Successive measurements of particle positions were originally considered in the context of interference of two independent BECs to mimic a simultaneous measurement of particle positions~\cite{Javanainen1996}.
The result by Syrwid and Sacha indicates that an yrast state should be interpreted as a state in which dark solitons are present but their positions are uncertain.
Although the expectation value of $\hat{\psi}^\dagger(x)\hat{\psi}(x)$ in a single yrast state is uniform, a single simultaneous measurement of all the positions yields a soliton density profile.

By using Eq.~(\ref{eq:soliton_wave}), it can be shown that if the quantum soliton state $\ket{N,X;P}$ is chosen as an initial state, the state after $N-N'$ measurements is identical to $\ket{N',X;P}$ with probability one.
In other words, the sequence of $\{\ket{N',X;P}\}_{N'=1}^N$ is an exact solution of the measurement dynamics.

In the free-boson limit, $c=0$, the relation is clearer; it can be analytically shown that a quantum state obtained after $N-N'$ measurements of particle positions on an yrast state $\ket{N,P}_\mathrm{yr}$ is given by a quantum soliton state $\ket{N',X;P}$ with probability very close to one when $N\gg N'$, where the position of a dark soliton $X$ is random and strongly depends on the realization of measurement outcomes~\footnote{E. Kaminishi, T. Mori, and S. Miyashita, in preparation}.
However, it is still open to prove the corresponding result for a small but finite coupling $c>0$.

\begin{acknowledgments}
The present work was supported by JSPS KAKENHI Grants No. JP16J03140 and Grants-in-Aid for Scientific Research C (No. 17K05508, No. 18K03444) from MEXT of Japan.
\end{acknowledgments}

\bibliography{darksoliton}

\end{document}